\documentclass[aps,prl,epsfigure,showpacs,twocolumn]{revtex4}
\usepackage{amsmath}
\usepackage{amstext}
\usepackage{latexsym}
\usepackage{psfig}
\usepackage{graphicx}
\usepackage{amsfonts}
\usepackage{amssymb}

\newcommand{\ket}[1]{\left\vert#1\right\rangle}

\newcommand{\pro}[2]{\left\vert#1\rangle\langle#2\right\vert}

\newcommand{\bra}[1]{\left\langle#1\right\vert}

\newcommand{\sgn}{\text{sgn}} 
\begin{document}

\title{Quantum state transfer in imperfect artificial spin networks}
\author{M. Paternostro$^{1}$, G. M. Palma$^{2}$, M. S. Kim$^{1}$, and G. Falci$^{3}$}
\affiliation{$^{1}$School of Mathematics and Physicis, The Queen's University, Belfast, BT7 1NN, UK\\
$^{2}$NEST-INFM \& Dipartimento di Tecnologie dell'Informazione,
Universita' di Milano, Via Bramante 65,
26013 Crema, Italy\\
$^{3}$MATIS-INFM \& Dipartimento di Metodologie Fisiche e Chimiche,
Universita' di Catania, Viale A. Doria 6, 95125 Catania, Italy}
\date{\today}

\begin{abstract}
High-fidelity quantum computation and quantum state transfer are possible in short spin chains. We exploit a system based on a dispersive qubit-boson interaction to mimic $XY$ coupling. In this model, the usually assumed nearest-neighbors coupling is no more valid: all the qubits are mutually coupled. We analyze the performances of our model for quantum state transfer showing how pre-engineered coupling rates allow for nearly optimal state transfer. We address a setup of superconducting qubits coupled to a microstrip cavity in which our analysis may be applied. 
\end{abstract}
\maketitle


Many protocols for Quantum Information Processing (QIP) assume an arbitrary amount of control over the evolving system. However, any external influence may introduce errors and decoherence. Moreover, having a fine single-qubit control over a large register may be hard, especially for closely spaced subsystems. This has motivated some recent proposals in order to reduce the amount of the control over a quantum computer~\cite{janzing}. A register can be controlled by global action on clusters of qubits or via operations on just an ancillary {\it control} qubit that is programmable to any desired dynamics~\cite{janzing}. Another possibility is the engineering of an interaction able to accomplish a prefixed task. Thus, a process is formulated according to the program: initialization of the register, evolution by the designed interaction and measurement without further interference on the dynamics. 

Recently, it has been recognized that Heisenberg and $XY$ interactions with nearest-neighbor (NN) couplings can be used to transfer a quantum state through a network of qubits virtually without external control~\cite{boseekert}. For $XY$ couplings, linear chains of three NN coupled qubits achieve a perfect transfer fidelity and can be used as building blocks for a longer communication wire. This strategy is useful in those situations where the use of a photonic bus (conventionally accepted as a good information-carrier) is not easy or convenient. Moreover, a three-qubit chain of its $XY$ coupling allows for universal quantum computation~\cite{boseleung}. Very recently, quantum cloning in many-qubit networks has been studied~\cite{massimoclone}. 

The appealing possibilities offered by even simple networks motivate the research for practical systems in which $XY$ interactions with possibly tunable couplings can be realized. We use the off-resonant interaction of a group of qubits with a common bus in order to simulate spin networks of tunable $XY$ couplings. 
In this model, the NN restriction is naturally relaxed and we study the efficiency of a state transfer process. The insight we gain through this kind of simulation under controllable conditions can then be applied to real systems of solid-state physics where the amount of control is in general smaller. Properly designing the coupling rates, we show that the transfer fidelity in chains of different lengths can be nearly optimal. Furthermore, in proper conditions, once we set the strengths of the inter-qubit couplings in a chain of $N$ elements, the quantum state to be transferred can be collected at specific times at any of the $N-k$ qubits of the chain ($k=0,1,..,N-2$). This may represent an advantage with respect to strategies that bypass the chain and connects directly the sender qubit to the receiver. In the latter, the couplings have to be re-designed each time the receiver's position is changed. 

A setup is addressed combining quantum optics and superconducting quantum interference devices (SQUIDs)~\cite{shon} that can embody our model. This choice is motivated by the advantages of a strong coupling regime,
 the fixed positions of the qubits in the cavity and the long life-times achievable in a high-quality factor (high-$Q$) cavity. These features have been recently exploited for solid state-quantum optics interfaces~\cite{noi}.  


{\it The system- }We consider $N\ge2$ qubits placed inside a cavity providing a single boson mode. We assume that, via  an external potential $\Phi$, we can modulate the transition energy $E_{q,i}$ of the qubits. This is possible in many QIP devices such as trapped ions, neutral atoms (through Stark fields) and SQUID-based systems (via a magnetic flux modulating the Josephson energy~\cite{shon}). 

The free Hamiltonian of the qubit system is $\sum^{N}_{i=1}\hat{H}_{i}=(1/2)\sum^{N}_{i=1}E_{q,i}(\Phi)\hat{\sigma}^{z}_{i}$  where $\hat{\sigma}^{z}_{i}$ is the $z$-Pauli operator of the $i^{th}$ qubit. Its eigenstates $\{\ket{\pm}\}_{i}$ are the basis for the $i^{th}$ qubit. The dependence of the qubit energies on $\Phi$ is explicitly shown. In many cases, the single addressing of the qubits in a register is a difficult task. We thus assume that $\Phi$ acts collectively on the qubits that are closely spaced in the cavity. The cavity field mode of its frequency $\omega_{a}$, described by the annihilation (creation) operator $\hat{a}$ ($\hat{a}^{\dagger}$), is coupled to the qubits. The wavelength $\lambda$ of the field is taken much longer than the dimension $d$ of each qubit and their separation so that any dependence on the position in the cavity is neglected. This assumption can be relaxed if necessary. We consider the generally valid field-qubit interaction model  
\begin{equation}
\label{Ham}
\sum^{N}_{i=1}\hat{H}_{i,a}=\sum_{i}\Omega_{i}(\hat{a}^{\dagger}+\hat{a})(\hat{\sigma}^{-}_{i}+\hat{\sigma}^{+}_{i})\hspace{0.5cm}{(\hbar=1)},
\end{equation}
where we assume the Rabi frequency $\Omega_{i}$ depends on a dimensionless parameter $\eta_{i}$ which can be designed qubit by qubit and set once for all. This is in the spirit of programming the system to accomplish a given task without interferences to its evolution. We have introduced the operators $\hat{\sigma}^{+}_{i}=(\hat{\sigma}^{-}_{i})^{\dag}=\ket{+}_{i}\!\bra{-}$. Altogether, the dynamics of the system is given by $\hat{H}=\sum^{N}_{i=1}\hat{H}_{i}+\hat{H}_{a}+\sum^{N}_{i=1}\hat{H}_{i,a}$,
with $\hat{H}_{a}=\omega_{a}\hat{a}^{\dagger}\hat{a}$. 
Motivated by the recent achievement of high-$Q$ cavities (strip-line cavities with $Q\ge{10}^4$ are relevant to this work) and strong-coupling regime in some systems that can embody our model~\cite{noi,schoelkopf}, we start without considering dissipation. The main sources of decoherence in our proposal will be addressed later in this paper. 

We take $\Omega_{i}\ll\omega_{a},\,E_{q,i}$ under the rotating wave approximation. In the interaction picture and with the tunable detunings $\delta_{i}=\omega_{a}-E_{q,i}(\Phi)$, the Hamiltonian is
$\hat{H}=\sum^{N}_{i=1}\Omega_{i}\hat{a}^{\dag}\hat{\sigma}^{-}_{i}e^{i\delta_{i}t}+h.c.$
If $\delta_{i}\gg\Omega_{i}$, the qubits act as a dispersive intra-cavity medium that changes $\omega_{a}$ according to $\omega_{a}\rightarrow\omega_{a}+\sum_{i}{\Omega^{2}_{i}}/{\delta_{i}}$. No real energy-exchange is possible, in this case. This permits the elimination of the bosonic mode from the dynamics of the qubits. Now the evolution is ruled by an effective adiabatic Hamiltonian
 (where terms oscillating at frequency $\delta_{i}$ or higher are discarded) having the form of a generalized $XY$ model
\begin{equation}
\label{effettiva2}
\hat{H}_{e}\simeq\sum_{i<j}2x_{ij}\left(\hat{\sigma}^{+}_{i}\hat{\sigma}^{-}_{j}+\hat{\sigma}^{-}_{i}\hat{\sigma}^{+}_{j}\right)=\sum_{i<j}{x_{ij}}\left(\hat{\sigma}^{x}_{i}\hat{\sigma}^{x}_{j}+\hat{\sigma}^{y}_{i}\hat{\sigma}^{y}_{j}\right)
\end{equation} 
with $x_{ij}=\Omega_{i}\Omega_{j}/2\delta_{j}$ and the sum runs over the qubits having $\delta_{i}=\delta_{j}$. A state-independent term has been neglected~\cite{locale}. Similar adiabatic interactions were studied elsewhere~\cite{altri}. In particular Biswas and Agarwal cosidered a quantum state transfer in a particle-chain. For longer chains, however, their proposal turns out to be experimentally quite demanding~\cite{altri}.

Christandl {\it et al.}~\cite{boseekert} considered 
an $XY$ model with NN couplings showing that perfect state transfer is possible in chains of $2$ and $3$ qubits. However, Eq.~(\ref{effettiva2}) is about long-range, non-NN interactions and is a different model whose properties we study, here, with respect to the efficiency of quantum state transfer processes. 

{\it Engineered state transfer-} We denote $\ket{\underline{i}}=(\otimes^{N}_{k\neq{i}}\ket{-}_{k})\otimes\ket{+}_{i}$ for a state where only the $i^{th}$ qubit is in $\ket{+}_{i}$ and $\otimes$ denotes tensorial product. Because $[\hat{H}_{e},\sum^{N}_{i=1}\hat{\sigma}^{z}_{i}]=0$, $\hat{H}_{e}$ can be diagonalized in each subspace having an assigned number ${\ell}$ of excited qubits. In particular, the subspace with $\ell=0$ ({\it i.e.} all the qubits in the $\ket{-}$ state) is spanned by $\ket{\underline{0}}$. The dynamic of an arbitrary state with just one excited qubit is instead confined in the subspace with $\ell=1$, spanned by the orthonormal basis $\{\ket{\underline{i}}\}\,(i=1,..,N)$. With the decomposition $\hat{U}_{\ell}(t)\!=\!e^{-i\hat{H}_{e,{\ell}}t}\!=\!\sum^{n_{s}-1}_{k=0}e^{-i\epsilon^{k}_{\ell}t}\vert{\psi^{k}_{\ell}}\rangle\langle\psi^{k}_{\ell}\vert$ we calculate the evolved state of the network. Here, $\hat{H}_{e,\ell}$ is the restriction of Eq.~(\ref{effettiva2}) to the $n_{s}-$dimensional subspace with a given $\ell$ and $\epsilon^{k}_{\ell}$ is the eigenvalue  corresponding to the eigenstate $\ket{\psi^{k}_{\ell}}$. 

Let the initial state $\ket{in}$ of the network have the qubit labelled $1$ in a superposition of $\ket{-}_{1}$ and $\ket{+}_{1}$ while all the other qubits are in $\ket{-}$, {\it i.e.} $\ket{in}=\beta\ket{\underline{0}}+\gamma\ket{\underline{1}}$. 
This state evolves to
$\beta\ket{\underline{0}}+\gamma\sum^{N-1}_{k=0}e^{-i\epsilon^{k}_{1}t}p_{{k}\underline{1}}\ket{\psi^{k}_{1}}$,
with the projections $p_{{k}\underline{1}}=\bra{\psi^{k}_{1}}\underline{1}\rangle$. For quantum state transfer, we are interested in the transition amplitude $\bra{\underline{N}}\hat{U}_{1}(t)\ket{\underline{1}}=\sum^{N-1}_{k=0}e^{-i\epsilon^{k}_{1}t}p_{{k}\underline{1}}p^{*}_{{k}\underline{N}}$, whose square modulus gives the probability for the process $\ket{+}_{1}\rightarrow\ket{+}_{N}$ to occur. In general, the $N^{th}$ qubit at a time $t$ is in a mixed state $\rho_{N}=Tr_{12..N-1}(\hat{U}(t)\pro{in}{in}\hat{U}^{\dagger}(t))$.
The fidelity of the process is defined as the quantity ${\cal F}(\beta,\gamma)=(\bra{-}\beta^{*}+\bra{+}\gamma^{*})\rho_{N}(\beta\ket{-}+\gamma\ket{+})$. To cancel dependences on the initial state, we average ${\cal F}(\beta,\gamma)$ over the surface of the Bloch sphere as $\bar{\cal F}=(1/4\pi)\int{\cal F}(\beta,\gamma){d}\Sigma$ with $d\Sigma$ the surface element. 
\begin{figure}[ht]
\centerline{{\bf (a)}\hskip1.4cm{\bf (b)}\hskip1.7cm{\bf (c)}\hskip1.4cm{\bf (d)}}
\centerline{\psfig{figure=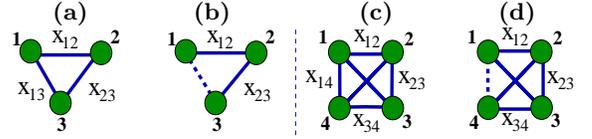,width=7.5cm,height=1.5cm}}
\caption{Coupling configuration for $N=3,4$. If the qubits are connected by all equally coupled, the systems are equivalent to triangular and squared clusters. If one connection is broken by reducing its coupling rate, the system becomes topologically equivalent to a two-qubit chain {\bf (b)} or, generally, to a superposition of inequivalent chains {\bf (d)}.}
\label{schema}
\end{figure}

\begin{figure} [bt]
{\bf (a)}\hskip3cm{\bf (b)}
\centerline{\psfig{figure=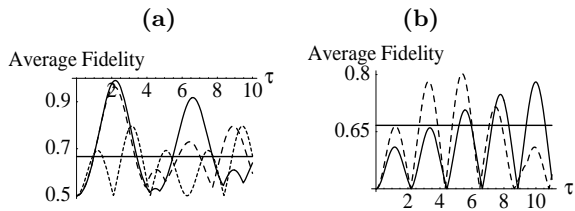,width=8.0cm,height=2.5cm}}
\caption{{\bf (a)}: $\bar{\cal F}$ vs. the rescaled time $\tau$ for $N=3$. The straight line is the bound for classical transfer. We show $\bar{\cal F}$ for the {\it all-equal couplings} case (dotted line) and the {\it engineered} cases with $f=10$ (solid line), $f=5$ (dashed line). {\bf (b)}: $\bar{\cal F}$ vs. $\tau$ for the $1\rightarrow{2}$ transfer process for $f=1.1$ (dotted line) and $f=5$ (solid line).}
\label{secca}
\end{figure}

We start with a simple case that captures the spirit of this study. The topology of a network depends crucially on the configuration of the couplings $x_{ij}$ in Eq.~(\ref{effettiva2}). In general, for equal $\delta_{i}$'s and $\Omega_{i}$'s, the coupled qubits form a graph of connected vertices. In Fig.~\ref{schema} we show the cases of $N=3,\,4$ that result in a triangular and a squared cluster respectively (Fig.~\ref{schema} {\bf (a)} and {\bf (c)}). Each vertex and solid line represent a qubit and a non-zero coupling, respectively. The natural question raised here is the performance of the state transfer in this network of connections. To analyze it, we explicitly solve the problem formulated above, considering just the relevant subspace with $\ell=1$ but for a generic number of qubits. For simplicity, we set $x_{ij}=x$. We note $n_{s}=N$ and find that $\hat{H}_{e,1}$ admits $N-1$ degenerate eigenvalues $\epsilon^{k}_{1}=-x$ ($k=0,..,N-2$) and the eigenvalue $\epsilon^{N-1}_{1}=(N-1)x$ completes the spectrum. The degenerate eigenspace is diagonalizable and an orthogonal basis can be built through the Gram-Schmidt algorithm. We get $\hat{U}_{1}(t)=e^{ixt}\sum^{N-2}_{k=0}\ket{\psi^{k}_{1}}\bra{\psi^{k}_{1}}+e^{-i(N-1)xt}\ket{\psi^{N-1}_{1}}\bra{\psi^{N-1}_{1}}$ with $\ket{\psi^{N-1}_{1}}=(1/\sqrt{N})\sum^{N}_{i=1}\ket{\underline{i}}$. The transfer probability is then $\vert\bra{\underline{N}}{\hat{U}_{1}(\tau)}\ket{\underline{1}}\vert^2=(2/N^2)\left[1-\cos(N\tau)\right]$, where $\tau=xt$ is the rescaled interaction time. The resulting average fidelity $\bar{\cal F}$ is plotted against $\tau$ in Fig.~\ref{secca} (dotted line). The maximum of this function has to be contrasted with $2/3$, the best fidelity achievable for the transfer of a qubit state through a classical channel~\cite{horodecki}. It is clear that $\bar{\cal F}$ may be higher than the classical limit.
However, trying to extend the system, we find that already for $N=4$ $\bar{\cal F}_{max}\simeq{2/3}$ and the quantum channel becomes useless. 

For $N=3$, the unwanted coupling $x_{13}$ does not compromise the state transfer, even if the fidelity is not optimal. However, a strategy to enlarge the range of $N$ for which this quantum process is still worthy is desirable. An intuitive approach is to get rid of the redundant connections. For example, as shown in Figs.~\ref{schema} {\bf (b)} and {\bf (d)}, by cutting the coupling $1\leftrightarrow{N}$, the clusters become equivalent to a chain of three qubits {\bf (b)} and to a superposition of elementary chains (three inequivalent paths connecting $1$ to $4$ can be found in Fig.~\ref{schema} {\bf (d)}).
 The cut may be realized by properly setting {\it ab initio} the Rabi frequencies $\Omega_{j}$'s. This is possible by an appropriate choice of each $\eta_{i}$ defined to be related to $\Omega_{i}$ in Eq.~(\ref{Ham}). For $N=3$, for example, we take $\Omega_{1}=\Omega_{3}\ll\Omega_{2}$ to get $x_{12}=x_{23}\gg{x}_{13}$. This reduces the complexity of the network to a $3$-qubit chain. We set $x_{12,23}=x$ and $x_{13}=x/f$, with $f>1$. The decomposition of $\hat{H}_{e,1}$ can be found analytically with the eigenvalues $\epsilon^{0}_{1}=-{x}/{f},\,\epsilon^{1,2}_{1}=(x/2f)(1\mp\sqrt{1+8f^2})$.
They correspond to the states 
$\ket{\psi^{0}_{1}}=({1}/{\sqrt{2}})(-\ket{\underline{1}}+\ket{\underline{3}})$ and $\vert{\psi^{1,2}_{1}}\rangle={\cal N}_{1,2}(\ket{\underline{1}}-({\epsilon^{2,1}_{1}}/{x})\ket{\underline{2}}+\ket{\underline{3}})$
with the normalizations ${\cal N}_{1,2}=x[2x^2+(\epsilon^{2,1}_{1})^2]^{-1/2}$. We find $\bra{\underline{3}}\hat{U}_{1}(t)\ket{\underline{1}}=-{(1/2)}e^{i{x}t/f}+{\cal N}^{2}_{1}e^{-i\epsilon^{1}_{1}t}+{\cal N}^{2}_{2}e^{-i\epsilon^{2}_{1}t}$ that reveals the competition between the different paths the system can follow from the sender to the receiver: the path connecting $1$ to $3$ via $x_{13}$ and the one through $\ket{\underline{2}}\propto{\cal N}^{-1}_{2}\ket{\psi^2_{1}}-{\cal N}^{-1}_{1}\ket{\psi^1_{1}}$. The transfer fidelity ${\cal F}(0,1)\equiv\vert\langle\underline{3}\vert\hat{U}_{1}(t)\vert\underline{1}\rangle\vert^2$ resulting from this interference effect is 
\begin{equation}
\label{fedeltagenerale}
{\cal F}(0,1)=\frac{1}{4}+2{\cal N}^{2}_{1}{\cal N}^{2}_{2}\cos(\Delta\epsilon^{21}_{1}t)
+\sum^{2}_{i=1}{\cal N}^{4}_{i}-{\cal N}^{2}_{i}\!\cos(\Delta\epsilon^{i0}_{1}t)
\end{equation}
with $\Delta\epsilon^{ij}_{1}=\epsilon^{i}_{1}-\epsilon^{j}_{1}$ ($i,j=0,1,2$).
Eq.~(\ref{fedeltagenerale})
reduces to what is found by Christandl {\it et al.}~\cite{boseekert} when $f\rightarrow{\infty}$. 
For this coupling-engineered system, it is ${\cal F}(0,1)\simeq0.973$ for $f=10$ and ${\cal F}(0,1)\simeq0.898$ for $f=5$ at $\tau\simeq{2f\pi/\sqrt{1+8f^2}}$. As time goes by, the interferences lead to collapses and revivals of the fidelity. The average fidelity $\bar{\cal F}$ can be computed and it is shown in Fig.~\ref{secca} {\bf (a)} where it is seen that the classical bound can be beaten. 

\begin{figure}[t]
{{\bf (a)}\hskip4.5cm{\bf (b)}}
\centerline{\psfig{figure=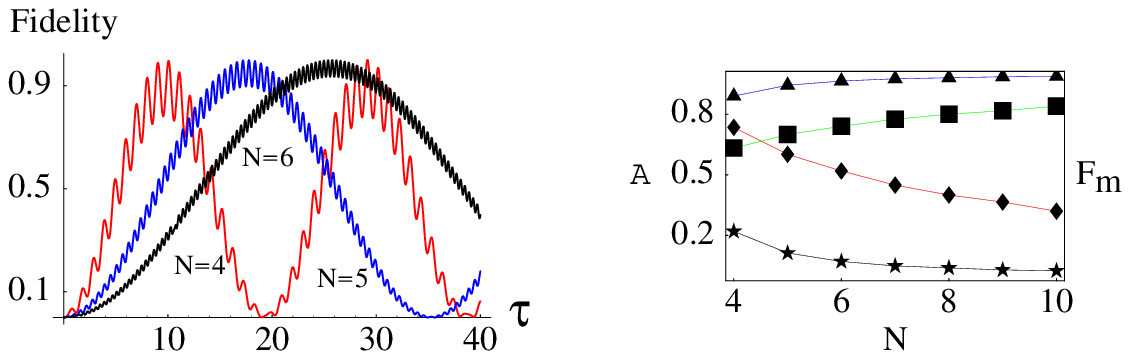,width=9.1cm,height=3.0cm}}
{{\bf (c)}\hskip4.5cm{\bf (d)}}
\centerline{\psfig{figure=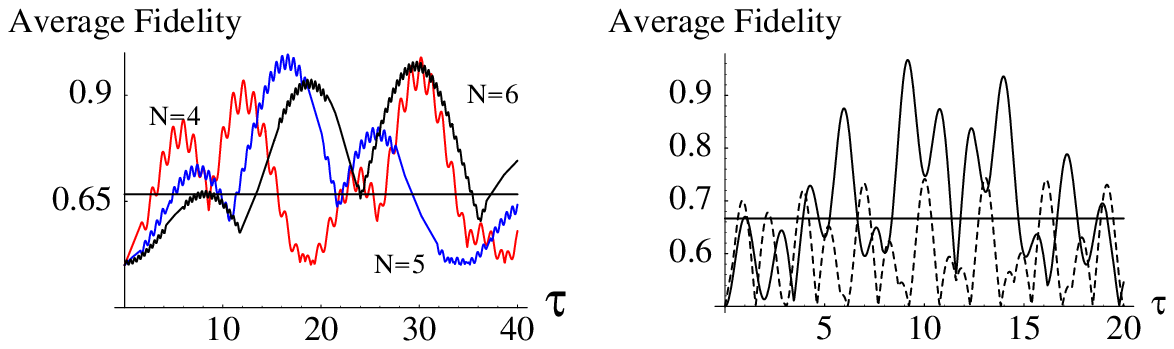,width=9.1cm,height=3.0cm}}
\caption{${\bf (a)}$: Fidelity ${\cal F}(0,1)$ vs. $\tau$ for $N=4,5,6$ and $f=5$. ${\bf (b)}$: Amplitude $A$ of the fast oscillations of ${\cal F}(0,1)$ potted against ${N}$ (left vertical axis) for $f=1.1$ ($\blacklozenge$) and $f=5$ ($\bigstar$). The right vertical axis shows $F_{m}$ vs. $N$ for $f=1.1$ ($\blacksquare$) and $f=5$ ($\blacktriangle$). {\bf (c)}: Average fidelity $\bar{\cal F}$ for the values in {\bf (a)}. {\bf (d)}: $\bar{\cal F}$ of the ``$1\rightarrow4$'' (solid line) and "$1\rightarrow3$"  (dotted line) process for $N=4$ and $f=1.1$.}
\label{insieme}
\end{figure}

Increasing the dimension of the network the diagonalization of Eq.~(\ref{effettiva2}) becomes demanding due to the dimension of the subspace with $\ell=1$. However, a recurrence law in $f$ and $N$ for the spectrum of $\hat{H}_{e}$ can be found to allow an analytical expression for the transfer fidelity regardless of the length of the cluster. When $x_{1N}=x/f$ is the smallest coupling, we find that $\hat{H}_{e}$ has $N-3$ degenerate eigenvalues, the rest of the spectrum being given by ${\epsilon^{N-3}_{1}=-x/f}$ and
$\epsilon^{N-k}_{1}=\frac{1+(N-3)f^2}{2f}\left\{1+\sgn{(k)}\sqrt{1+\frac{4(N-1)f^2}{[1+(N-3)f^2]^2}}\right\}x$
for $k=1,2$ and $\sgn(1,2)=\pm$. 
The state $\ket{\underline{N}}$ does not belong to the subspace spanned by the degenerate eigenstates. The decomposition of $\hat{U}_{1}$ can thus be effectively performed just considering the eigenstates $\ket{\psi^{N-3}_{1}}=({1}/{\sqrt 2})\left(-\ket{\underline{1}}+\ket{\underline{N}}\right)$
and
$\ket{\psi^{N-k}_{1}}={\cal N}_{N-k}\left[\ket{\underline{1}}+\ket{\underline{N}}+\frac{({\epsilon^{N-3}_{1}+\epsilon^{N-k}_{1}})}{x({N-2})}\sum^{N-1}_{i=2}\ket{\underline{i}}\right]$,
 where ${\cal N}_{N-k}=x(N-2)[2(N-2)^2x^2+(\epsilon^{N-3}_{1}+\epsilon^{N-k}_{1})^2(N-2)]^{-1/2}$. The fidelity keeps the structure 
of Eq.~(\ref{fedeltagenerale}) with $\epsilon^{k}_{1}\rightarrow\epsilon^{N-k}_{1}~(k=1,2)$, $\epsilon^{0}_{1}\rightarrow\epsilon^{N-3}_{1}$. As $f$ increases ${\cal N}_{N-1}\rightarrow{0},\,{\cal N}_{N-2}\rightarrow{1/\sqrt{2}}$ and an $N$-dependent value of $\tau$ when ${\cal F}(0,1)\simeq{1}$ can be found. This is shown for $N=4,5,6$ in Fig.~\ref{insieme}, where $f=5$ has been taken. The plot shows the fast oscillations of ${\cal F}(0,1)$ around an average that is sinusoidal with $\tau$. Practically, this may be a problem as the value of the fidelity depends on the ability of stopping the process at precise moments. The instantaneous value of ${\cal F}(0,1)$ thus loses significance in favour of a mean value ${F}_{m}$ defined as the semi-distance between the maximum and the minimum of the oscillations. Because of the decreasing contribution by the {\it beating} term $\propto\cos[\Delta\epsilon^{N-1,N-2}_{1}t]$, the amplitude $A$ of the fast oscillations decreases as $N$ and $f$ grow. This stabilizes the fidelity around the slower behavior.
A compromise between a fast process and the uncertainty in ${\cal F}(0,1)$ is desirable. From Fig.~\ref{insieme} ${\bf (b)}$ ($\bigstar$ and $\blacktriangle$) we see that $f=5$ is a suitable choice, allowing a high fidelity (${F}_{m}\,{\gtrsim}\,0.9$ for $N\ge4$) within $\tau\lesssim60$ (up to $N=10$) and nearly optimal average fidelity $\bar{\cal F}$, as shown in Fig.~\ref{insieme} ${\bf (c)}$. However for $f=1.1$ and $4<N<10$ we find $F_{m}\gtrsim0.7$ (Fig.~\ref{insieme} {\bf(b)}, $\blacksquare$) and $\bar{\cal F}\gtrsim0.9$. 
It is noticeable that, even for values of $f$ that do not optimize the above analysis, we still get very good values of $\bar{\cal F}$. Furthermore, we stress that, once we set the $x_{ij}$'s allowing for the "$1\rightarrow{N}$" transfer, the same network can be used, in proper conditions, for "$1\rightarrow{j}$" processes too ($2\le{j}\le{N}-1$). This is shown in Fig.~\ref{secca} {\bf (b)} for $N=3$ and and in Fig.~\ref{insieme} {\bf (d)} for $N=4$.

{\it The proposed setup- }As a setup combining strong coupling regime and fixed qubits positions in the cavity, we address a cavity-superconducting qubits system~\cite{noi,schoelkopf}. 
An array of mutually coupled Josephson junctions may be used as a high-fidelity quantum channel~\cite{fazio}. Our approach is different as we exploit the advantages of a dispersive bus. 
We take $N$ SQUIDs~\cite{shon} in a $1D$ superconducting strip-line resonator ($Q\gtrsim10^4$, $\omega_{a}\simeq10\,$GHz, $\lambda\simeq{1}\mbox{cm}\gg{d}\simeq{1}\,\mu$m). The whole setup could be fabricated via nanolithographic techniques allowing for a precise control and calibration of the characteristics of the system. 
The strip-line cavity minimizes the photon-losses and protects, to some extent, the qubits from the environment. The dephasing due to charge-coupled, low-frequency noise is a dangerous source of decoherence~\cite{elisabetta}. However, at the {\it degeneracy point} of the SQUIDs~\cite{shon,noi}, each qubit is encoded in the space spanned by the equally-charged states $\ket{\pm}_{i}=(\/\sqrt 2)(\ket{0}\pm\ket{2e})$ ($2e$ is the charge of a Cooper pair). The environment is not able to distinguish between these states which are less sensitive to external charge fluctuations~\cite{francescopino,pinoproc}. Some estimates puts the dephasing rate in the range of $1\,\mu{sec}$~\cite{pinoproc}, in a resonant single SQUID-cavity systems. On the other hand, for $\Omega_{i}\sim{100}$MHz and $\delta_{i}\sim1$GHz, the Purcell effect in this off-resonant setup enhances the characteristic life-times up to $\simeq50\,\mu{sec}$, suitable for quantum state transfer up to $N=10$ ($f=5$). We have solved the master equation describing the evolution of networks of up to $5$ qubits, when the decay of cavity field and qubits is included. We found that the resulting dynamics follow the trends shown in Fig.~\ref{insieme} ${\bf (a)}$ without significant deviations. In this setup, the qubits energies $E_{q,i}$ are the Josephson coupling energies and the control $\Phi$ is a proper magnetic flux piercing the {\it whole} group of SQUIDs. Small relative differences between the detunings (up to $1\,$MHz) do not affect the transfer fidelity. In refs.~\cite{noi,schoelkopf}, qubits and cavity mode are capacitively coupled and the parameters $\eta_{i}$ that can be set properly designing, at the building stage, the capacities between the qubits and the the cavity. 
The initialization of the system can be performed if two values of $\Phi$ can be arranged. For $\Phi=\Phi_{1}$, we assume that the $x_{ij}$'s are small enough to turn off the mutual couplings. At low temperatures (assumed to guarantee the charging regime~\cite{shon}), the off-resonant coupling to the cavity makes the probability that a qubit is in $\ket{+}_{i}$ negligible. An electrode coupled to qubit $1$, then, provides a pulse that prepares the desired state. Switching to the proper $\Phi_{2}\ll\Phi_{1}$, suitable for state transfer, the process begins. Setting back $\Phi=\Phi_{1}$, the interaction can be stopped.

{\it Remarks- }A dispersive qubits-boson interaction is able to mimic an $XY$ model with tunable couplings. However, unwanted non-NN connections appear, complicating the dynamics of the network. We have studied quantum state transfer in this generalized $XY$ model, showing that the spoiling effect of the redundant connections can be bypassed. We have considered a setup combining quantum optics and SQUIDs, identifying the parameters discussed in the theoretical model and addressing the suitable strategies for quantum state transfer.

{\it Acknowledgements- }We acknowledge support by UK EPSRC, Korea Research Foundation (2003-070-C00024) and IRCEP. We thank R. Fazio for discussions.


\begin{thebibliography}{99}
\bibitem{locale} One-qubit terms are usually assumed to be compensated.
\end{thebibliography}
\end{document}